\documentclass[iop]{emulateapj}
\shorttitle{PPISN model for PTF12dam}
\shortauthors{Tolstov et al.}

\usepackage{color}
\usepackage{enumerate}
\usepackage{amsmath}
\usepackage{textcomp}
\usepackage{longtable}
\usepackage{ulem}

\usepackage{scalefnt}
\newcommand\smaller[2][0.85]{{\scalefont{#1}#2}}

\begin{document}

\title{Pulsational Pair-Instability Model for Superluminous Supernova PTF12dam:
Interaction and Radioactive Decay}

\author{Alexey Tolstov\altaffilmark{1}, Ken'ichi Nomoto\altaffilmark{1,8}, Sergei Blinnikov\altaffilmark{2,3,1}, Elena Sorokina\altaffilmark{4}, Robert Quimby\altaffilmark{5,1}, Petr Baklanov\altaffilmark{2,6,7}}

\affil{\altaffilmark{1} Kavli Institute for the Physics and Mathematics of the Universe (WPI), The
University of Tokyo Institutes for Advanced Study, The University of Tokyo, 5-1-5 Kashiwanoha, Kashiwa, Chiba 277-8583, Japan} 

%\affil{\altaffilmark{2} Sede Andina, Universidad Nacional de Rio Negro, Mitre 630 (8400) Bariloche, Argentina}

\affil{\altaffilmark{2} Institute for Theoretical and Experimental Physics (ITEP), 117218 Moscow, Russia} 

\affil{\altaffilmark{3} All-Russia Research Institute of Automatics (VNIIA), 127055 Moscow, Russia}

\affil{\altaffilmark{4} Sternberg Astronomical Institute, M.V.Lomonosov Moscow State University, 119991 Moscow, Russia}

\affil{\altaffilmark{5} Department of Astronomy, San Diego State University, San Diego, CA 92182, USA}

\affil{\altaffilmark{6} Novosibirsk State University (NSU), Novosibirsk 630090, Russia}

\affil{\altaffilmark{7} National Research Nuclear University (MEPhI),  Moscow 115409, Russia}

\email{$^{*}$ E-mail: alexey.tolstov@ipmu.jp}

\submitted{Accepted for publication in the Astrophysical Journal}
\journalinfo{Accepted for publication in the Astrophysical Journal}
\slugcomment{Accepted for publication in the Astrophysical Journal on 28 Dec 2016}

\begin{abstract}
\noindent
Being a superluminous supernova (SLSN), PTF12dam can be explained by a $^{56}$Ni-powered model, a magnetar-powered model or an interaction model. We propose that PTF12dam is 
a pulsational pair instability supernova, where the outer envelope of a progenitor is ejected during the pulsations.
 Thus, it is powered by double energy source: radioactive decay of $^{56}$Ni and a radiative shock in a dense circumstellar medium. To describe multicolor light curves and spectra we use radiation hydrodynamics calculations of {\sc STELLA} code. We found that light curves are well described in the model with 40M$_{\odot}$ ejecta and 20-40M$_{\odot}$ circumstellar medium. The ejected $^{56}$Ni mass is about 6M$_{\odot}$ which results from explosive nucleosynthesis with large explosion energy (2-3)$\cdot$10$^{52}$ ergs. In comparison with alternative scenarios of pair-instability supernova and magnetar-powered supernova, in interaction model all the observed main photometric characteristics are well reproduced: multicolor light curves, color temperatures, and photospheric velocities.
\end{abstract}


%=============================

\section{INTRODUCTION}
\label{sec:intro}
\noindent

\footnotetext[8]{Hamamatsu Professor.}

At the moment there is no universally accepted model for superluminous supernovae. Several scenarios are widely discussed \citep[see e.g.][for review]{Quimby2014}: the explosion of a star with a large initial mass
greater than 140M$_{\odot}$ (pair-instability supernova (PISN)) with the production of huge amount of radioactive nickel 
$M$($^{56}$Ni) up to 57M$_{\odot}$ \citep{HegerWoosley2002}; a spinning-down millisecond magnetar that transforms the rotational energy into the energy of the SN ejecta; or an interaction of the SN ejecta with the surrounding extended and dense circumstellar matter (CSM) that transforms the kinetic energy of the shock into radiation. In this paper we focus mostly on the interaction scenario.

When the ejecta interacts with CSM, the forward and reverse shocks merge into one dense shell. The interaction of the shell with CSM is generally considered as the most probable explanation for the high luminosity of Type II SLSNe (SLSN-II) \citep[e.g.,][]{Dessart2015}. But the shock interaction mechanism can also be applied for Type I SLSNe (SLSN-I) \citep{Sorokina2016}. 

PTF12dam is a SLSN-I which has spectra similar to SN 2007bi, but the peak luminosity is higher and estimated rise time is shorter. SN 2007bi was modelled in all three scenarios: PISN \citep{GalYam2009,Dessart2013,Kozyreva2014}, magnetar \citep{Nicholl2013} and interaction model \citep{Chatzopoulos2013}. Both magnetar and interaction models for SN 2007bi are based on  simple parametrizations and assumptions, and more accurate numerical simulations are required to analyse these scenarios. The PISN model naturally explains the long exponential tail of the light curve by radioactive decay of a large mass $^{56}$Ni, but magnetar model is more attractive alternative to reproduce the blue, weakly blanketed
and broad-lined spectra of SN 2007bi \citep{Dessart2012}. One more scenario was suggested by \citet{Moriya2010} in core-collapse supernova model with the mass of the radioactive nickel $^{56}$Ni about 6M$_{\odot}$ and explosion energy $E_{51}$ = $E$/10$^{51}$ erg = 36.

For PTF12dam \citet{Baklanov2015} have modeled this object as an explosion inside a CSM. The light curves are reproduced satisfactorily with a minimum set of model parameters and a modest explosion energy $E_{51}$ = 4. But the simulation was performed only up to +200 d after the luminosity peak. Later the observations revealed exponential decline of the light curve, which is not so abrupt, as we can expect from the interaction model. The power-law decline in the luminosity of PTF12dam continues at least for +400 days after the peak. To explain this behavior of the light curves, \citet{Chen2015} had to revise the magnetar model from the paper of \citet{Nicholl2013}, which gives overestimated values for the late points on the bolometric light curve. 

\begin{figure}
\includegraphics[width=80mm]{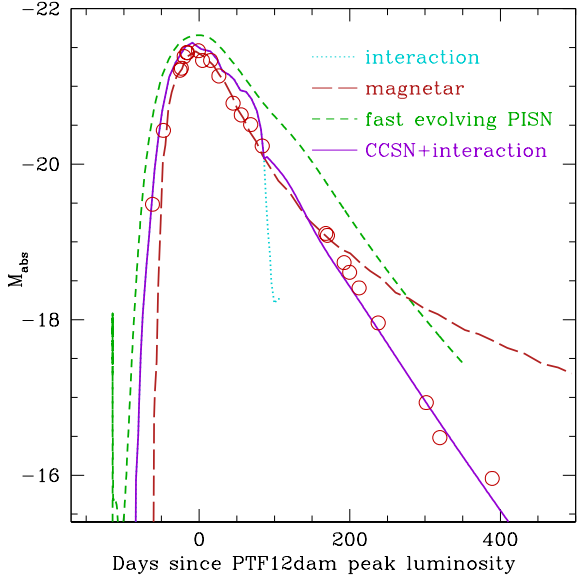}
\caption{Observed bolometric light curve of PTF12dam \citep{Chen2015} and a comparison of calculated bolometric light curves in various models: fast evolving PISN \citep[][P250 model]{Kozyreva2017}, interaction \citep{Baklanov2015} (quasi-bolometric light curve for the helium M53He48e40 model), magnetar \citep{Nicholl2013}, and the resulting bolometric light curve in two source model summing up the radiation of the interaction model \citep{Baklanov2015} and CCSN model  \citep{Moriya2010}.}
\label{slCmp}
\end{figure}

But the most natural explanation for the long tail of the light curve is the radioactive decay of $^{56}$Ni. The main idea described in this paper is to combine two sources of radiation: the shock interaction with CSM and radioactive decay of $^{56}$Ni. In Figure \ref{slCmp} we estimate the resulting bolometric light curve in this two source model summing up the radiation of the interaction and CCSN models constructed by \citet{Baklanov2015} and \citet{Moriya2010}, respectively. The result of our simplified procedures looks promising and self-consistent numerical simulations should clarify whether the interaction model is applicable for PTF12dam.

The bolometric light curve of PTF12dam could be also modelled by fast evolving PISN \citep{Kozyreva2017}. But in this paper in interaction model we consider the modeling of all the observed main photometric characteristics: multicolor light curves, color temperatures, and photospheric velocities.

The structure of the present paper is as follows. In section 2
we describe the models and the methods we use. In sections
3 we present the results of our calculations and
compare them with the observed broad band light curves
of PTF12dam. In the last section 4 we discuss the advantages and limitations of the interaction mechanism in comparison with PISNe and magnetar-powered SNe.

\section{MODELS AND METHODS}
\label{sec:models}
\subsection{Models}
\noindent

All presupernova models for this work are constructed from the core-collapse SN 2007bi model used in the paper of \citet{Moriya2010}. 

\begin{figure}
\includegraphics[width=80mm]{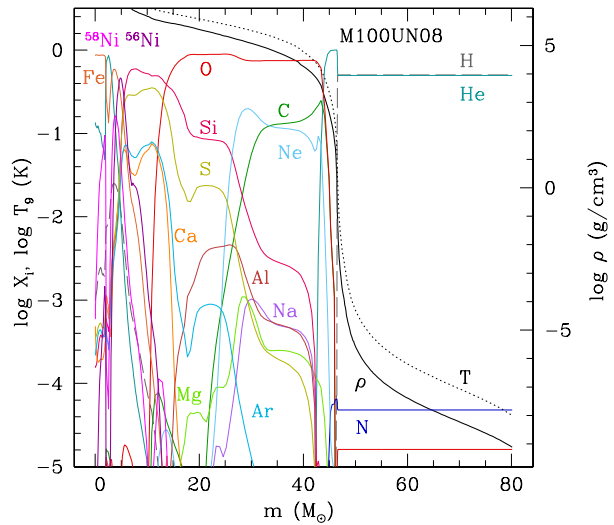}
\caption{Density, temperature and composition of the presupernova model \citep{UmedaNomoto2008}.}
\label{12damXprogenitor}
\end{figure}

\begin{figure}
\includegraphics[width=80mm]{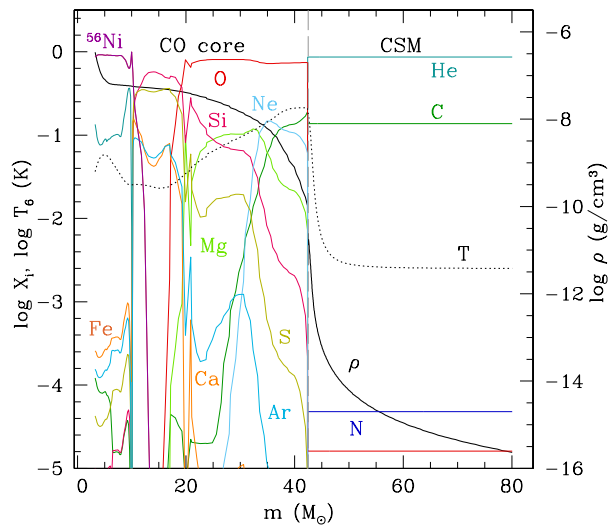}
\caption{Density and temperature of presupernova in the interaction model {\sc M80R165E20(CSM47)}, as well as its composition after explosive nucleosynthesis (1 day after explosion). The model is constructed by adding CSM with given composition to CO core of M100UN08 model. The boundary between the core and CSM at 43M$_{\odot}$ is shown by gray dashed line.}
\label{12damX}
\end{figure}

The progenitor (Figure \ref{12damXprogenitor}) has a main sequence mass of 100 M$\odot$ and the
metallicity of $Z = Z_\odot/200$ \citep{UmedaNomoto2008}, which is small enough
to avoid a large amount of wind mass loss.  The star undergoes strong pulsations at the end of Si-burning and evolves through the Fe core collapse. At the collapse, the mass of the C+O core is 43 M$\odot$. 
Pulsational pair instability during Si-burning \citep{HegerWoosley2002,Ohkubo2009} may eject some materials of even the C+O layers before collapse \citep{Woosley2007,Yoshida2016}.
For the presupernova model, the C+O star model of 43 M$\odot$ is
constructed by removing the He and H-rich layers \citep{Moriya2010}.
At a large explosion energy, \citet{Moriya2010} obtain a very large amount of radioactive $^{56}$Ni of 6.1 M$\odot$.

%%============================= TableBegin

\begin{deluxetable*}{lccccccccc}
\tablecaption{Model parameters \label{model}}
%\tablecolumns{5}
\tablewidth{0pt}
\tablehead{
\colhead{Model} & 
\colhead{$\log R$} & 
\colhead{$M$} &
\colhead{$M_{CSM}$} &
\colhead{$p$} &
\colhead{$E$} &
\colhead{$M$($^{56}$Ni)} & 
\colhead{$T$} &  
\colhead{Composition} &
\colhead{$N_{r}$} 
\\
\colhead{} & 
\colhead{(cm)} & 
\colhead{($M_{\odot}$)} &
\colhead{($M_{\odot}$)} & 
\colhead{} &
\colhead{($E_{51}$)} &
\colhead{($M_{\odot}$)} & 
\colhead{(K)} & 
\colhead{} & 
\colhead{} 
}  

{\sc M68R158E20(CSM6)}  & 15.8 & 68 & 25 & 2.0 & 20 & 6 & 2500 & He:C=6:1 & 289
\\ 
{\sc M66R170E27(CSM19)} & 16.0 & 66 & 23 & 2.0 & 27 & 6 & 2500 & O:C=4:1  & 192
\\ 
{\sc M58R165E20(CSM37)} & 16.5 & 58 & 15 & 2.0 & 20 & 6 & 2500 & He:C=6:1  & 289
\\ 
{\sc M80R165E20(CSM47)} & 16.5 & 80 & 37 & 2.0 & 20 & 6 & 2500 & He:C=6:1 & 289
\\
{\sc M87R165E50(CSM70)} & 16.5 & 87 & 44 & 2.0 & 50 & 6 & 2500 & He  & 289
\\

\vspace{-0.2cm}

%\enddata
%\vspace{-0.8cm}
\tablecomments{The first column shows the model name and the index number of the model from Table \ref{modelTable2} (see Appendix) in brackets.  The numbers shown are the radius of the model, the total mass $M=M_{\rm ej}+M_{\rm CSM}$, the mass of CSM, the index of the power-law CSM density profile, the explosion energy, the mass of $^{56}$Ni, the temperature of the wind, the composition of the wind, and the number of radius zones. The mass of SN ejecta in all the models $M_{\rm ej}$=40$M_{\odot}$.}
%\end{longtable*}
\end{deluxetable*}
%============================= TableEnd

\citet{Moriya2010} calculated the explosion of the pre-SN C+O star with following calculation of post-processing explosive nucleosynthesis. Explosions are induced by a thermal bomb
and followed by a one-dimensional Lagrangian code. The mass cut between the ejecta and the compact remnant is set at 3M$\odot$, so that the
ejecta contains 6.1M$\odot$ of $^{56}$Ni, which turns out to be consistent with the bolometric LC of SN 2007bi. The dynamics of the ejecta is followed until 1 day after the explosion, when the expansion already becomes homologous ($r\propto v$).

To make an interaction model we surround the ejecta at 1 day after the explosion ($R_{\rm ej}=3350 $R$_{\odot}$) by a rather dense CSM with the mass $M_{\rm CSM}$ extended to the radius $R_{\rm CSM}$. For all our models the CSM is outer radius $R_{\rm CSM}$ of the CSM is about $10^6$ R$_{\odot}$, or $\sim$ 10$^{17}$ cm. 
We use power-law density distribution $\rho \varpropto r^{-p}$ for the CSM, which simulates the wind that surrounds the exploding star. For a steady wind, $p = 2$, but in the very last stages of the evolution of a presupernova star the wind may not be steady. For our models we varied $p$ in the range from $1.5$ to $3.5$. 

\begin{figure}
\includegraphics[width=80mm]{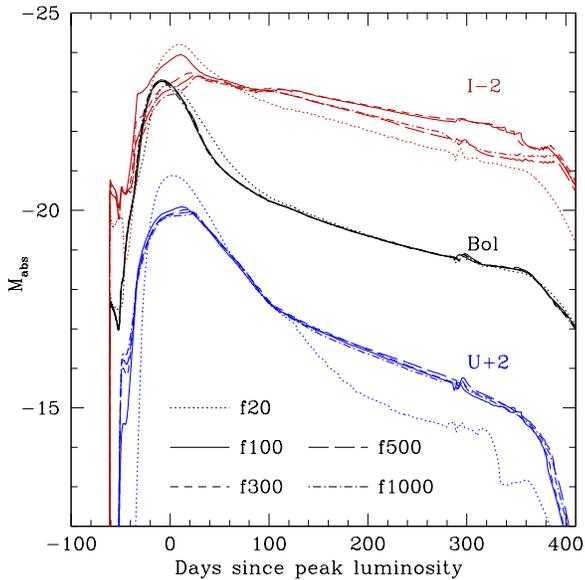}
\caption{Bolometric, U-band, and I-band light curves of the model {\sc M80R165E20(CSM47)} depending on the number of frequency groups. The number of radius zones for all the models $N_r=287$.}
\label{12damHighResFreq}
\end{figure}

\begin{figure}
\includegraphics[width=80mm]{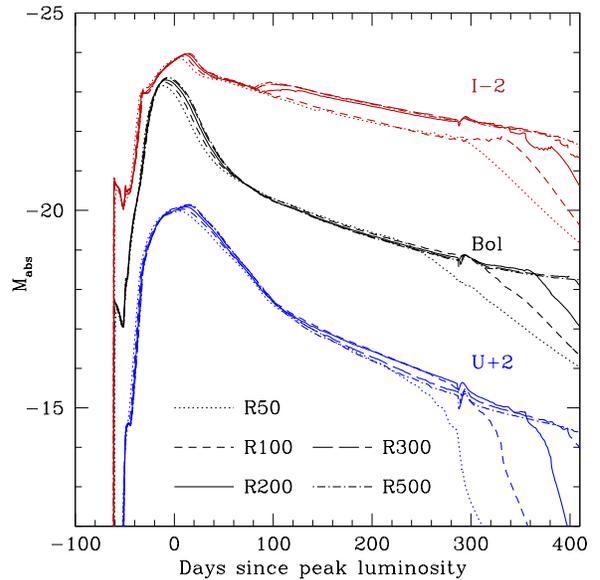}
\caption{Bolometric, U-band, and I-band light curves of the model {\sc M80R165E20(CSM47)} depending on the number of radius zones in CSM. The number of frequency groups for all the models $N_f=100$.}
\label{12damHighResR}
\end{figure}

Chemical elements in the \text{wind} are supposed to be distributed uniformly. Typically we use carbon-oxygen models with different C to O ratios or helium models (Figure \ref{12damX}). We also add some elements with higher atomic numbers (usually, 2\% of the total mass) with the abundances in solar proportion.
All models initially have $T = 2.5 \cdot 10^{3}$ K in the wind. Higher temperatures produce an artificial flash of light emitted by the huge CSM during its cooling
(Blinnikov \& Sorokina 2010).
The parameters of the most representative models for PTF12dam that are important for our discussion are shown in the Table \ref{model}.

\subsection{{\sc stella} Code}
\label{sec:stella}

For calculation of the light curves we use the multigroup radiation hydrodynamics numerical code {\sc stella} \citep{Blinnikov1998,Blinnikov2000,Blinnikov2006}. {\sc stella} solves implicitly time-dependent equations for the angular moments of intensity averaged over fixed frequency bands and computes variable Eddington factors that fully take into account scattering and redshifts for each frequency group in each mass zone. In our models we set $100$ frequency groups in the range from $1$ \AA\, to $5\times10^{4}$ \AA, and about $200$ radius zones for CSM. The ejecta has $97$ radius zones. In Figures \ref{12damHighResFreq} and \ref{12damHighResR} we checked the influence of the higher number of radius zones for CSM and frequency groups on the light curves. The frequency groups resolution higher than $100$ points can be important for red bands, and the radius zones resolution higher than $200$ points should be taken for +350d epoch. The high-resolution simulations are numerically expensive (more than $100$ hours per one run) to cover parameter space and we apply them only to check our best-fit models.

The explosion is initialized as a thermal bomb just above the mass cut, produce a shock wave that propagates outward. {\sc stella} run of initial model forms a shock wave at the border between the ejecta and the CSM. The shock converts the energy of the ordered motion of expanding gas to the thermal motion, which generates the emission.  The effect of line opacity is treated as an expansion opacity according to the prescription of \citet{EastmanPinto1993} (see also \citet{Blinnikov1998}). The opacity table includes $1.5 \times 10^5$ spectral lines from \citet{KuruczBell1995} and \citet{Verner1996}.

Modeling of near-infrared light curves in {\sc stella} is less reliable than optical light curves. {\sc stella} uses a rather poor line list in near infrared region, and taking into account of larger line list with millions of lines from large Kurucz table is now underway (E. Sorokina, private communication). 
%In this paper we limit our modeling of PTF12dam to optical passbands, which are the most reliable for {\sc stella} calculation.}

\section{RESULTS}
\label{sec:results}
\noindent
\subsection{Multicolor and bolometric light curves}

\begin{figure}
\includegraphics[width=80mm]{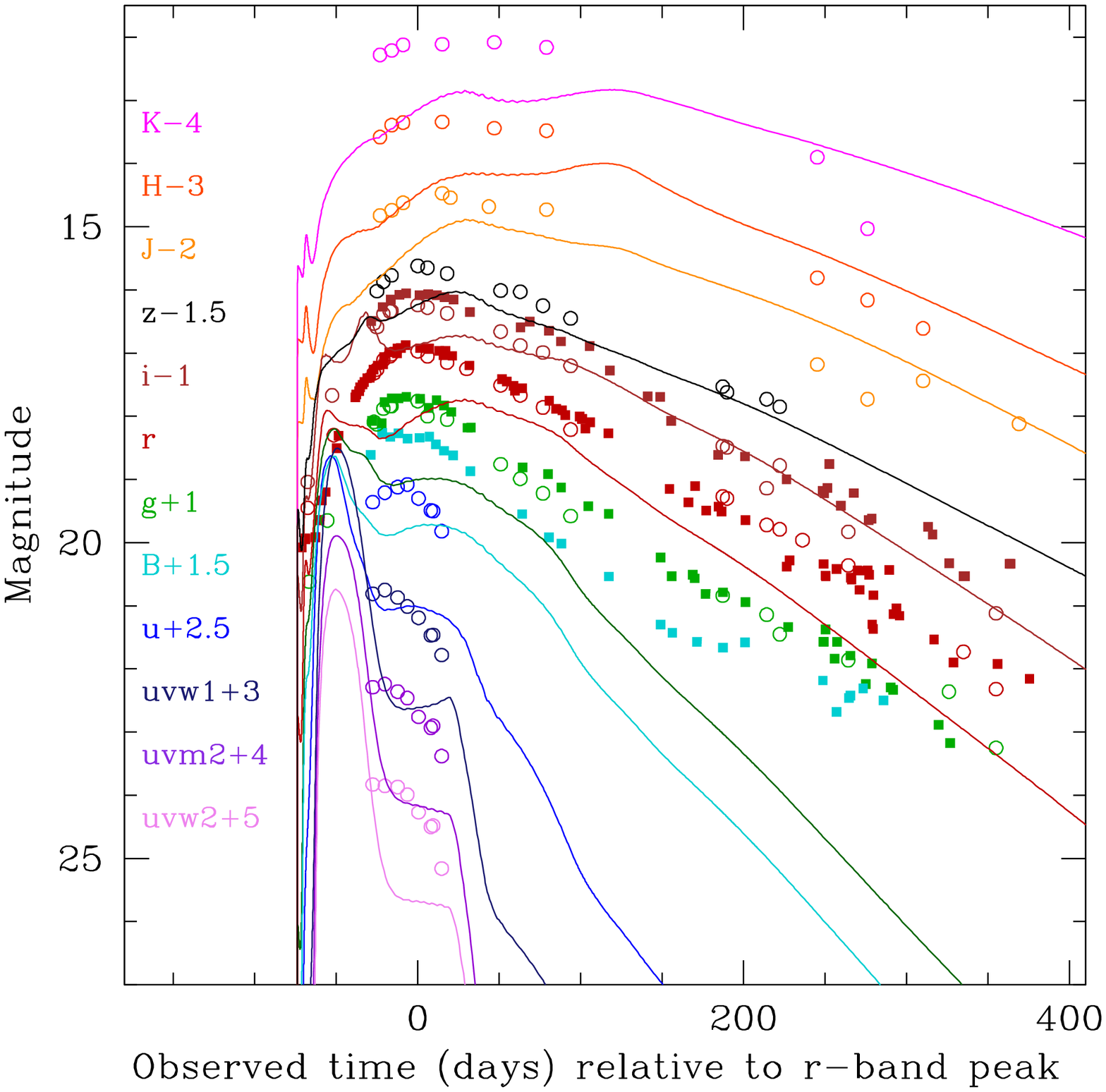}
\caption{Multicolor light curves simulation for PTF12dam in the model {\sc M66R170E27(CSM19)} and comparison with observations (filled 
squares from \citet{Vreeswijk2016}, open circles from \citet{Nicholl2013} and \citet{Chen2015}). Explosion time t$_{\rm 0}$=-74d. \\}
\label{mlcFirst}
\end{figure}

Figure \ref{mlcFirst} represents the result of the light curve calculations for the model \smaller{M66R170E27(CSM19)} that contains quite common feature of combined model: 2 peaks in the light curves in the
optical/UV wavelengths. The first peak in the luminosity is due to the interaction of the dense shell with the CSM and the second peak is due to the radioactive decay of $^{56}$Ni. 

The single peak in the observed light curve can be reproduced if the interaction peak has a longer duration and provides a smooth transition in the light curve from the interaction to the radioactive decay. The duration of the interaction phase is longer if the CSM is more massive.

In Figure \ref{mlcFirst} we also can see that the rise time of the modeled light curves is shorter than the observations. A longer rise time could be realized by changing chemical composition of the CSM. Model {\sc M66R170E27(CSM19)} has a carbon-oxygen CSM with the mass ratio C:O =1:4. Adding helium in the CSM the rise time of the light curve can be increased \citep{Sorokina2016}. At the temperature less than 11,000 K the opacity for the carbon-oxygen mixture at the same conditions is higher than for helium CSM. The shock wave heats up cold CSM and CSM opacity increases faster for CO mixture. The faster  increase of CSM opacity leads to the faster speed of the growth of the photospheric radius.

In fact, there are many parameters that could affect the light curve model: the explosion energy, the mass of the $^{56}$Ni, the density structure of the CSM, the radius of the CSM, and its composition. In calculations we cover only a part of the parameter space, choosing them from the most realistic physical conditions (see Appendix \ref{appModels} for the list of all the models).

The explosion energy $E$ lower than {$E_{51}<20$ can hardly provide a sufficient amount of $^{56}$Ni required for the luminous tail of the light curve \citep{UmedaNomoto2008}. The explosion energy $E_{51}>30$ foe is inconsistent with observations, because it forms too bright peak of the light curve due to increased reprocessing of the kinetic energy of the shock into radiation. 

A small mass of the CSM (5-10M$_{\odot}$) leads to a short duration of interaction phase (Figure \ref{mlcFirst}), while a large mass of the CSM ($>$ 40M$_{\odot}$) affects the $^{56}$Ni peak, making it too bright near $^{56}$Ni maximum. 

The multicolor light curves for the best-fit model are presented in Figure \ref{mlcBest}. The light curves at optical wavelengths are in good agreement with observations. Near-infrared wavelength light curves are fainter than observations and require more detailed opacity calculations, as we mentioned in section \ref{sec:stella}. The UV light curves are brighter in comparison with observations for all the models we considered. The UV emission can probably be extinct by a dust surrounding the supernova. Figure \ref{mlcBest} also shows that high-resolution simulations are preferable for r- and z-band near the luminosity peak. 

In the best-fit model we constructed helium-carbon CSM with mass $M$ $\sim$ 40M$_{\odot}$ and composition mass ratio He:C =9:1. The presence of carbon increases the opacity of the CSM by several orders of magnitude at temperature 7,000 K \citep{Sorokina2016}. The tail of g-band light curve in the models with pure helium CSM is too faint to fit the observational data.

\begin{figure}
\includegraphics[width=80mm]{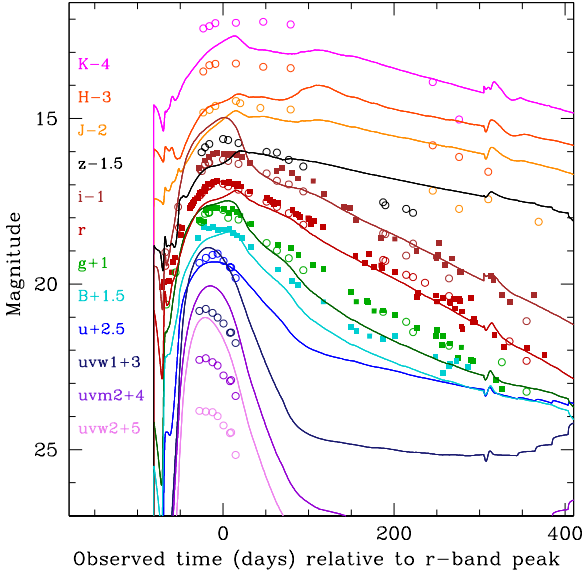}
\caption{Multicolor light curves simulation for PTF12dam in the model {\sc M80R165E20(CSM47)} with high-resolution simulation (N$_{\rm r}$=483, N$_{\rm f}$=500 and comparison with observations (filled 
squares from \citet{Vreeswijk2016}, open circles from \citet{Nicholl2013} and \citet{Chen2015}). Explosion time t$_{\rm 0}$=-82d.}
\label{mlcBest}
\end{figure}

\begin{figure}
\includegraphics[width=80mm]{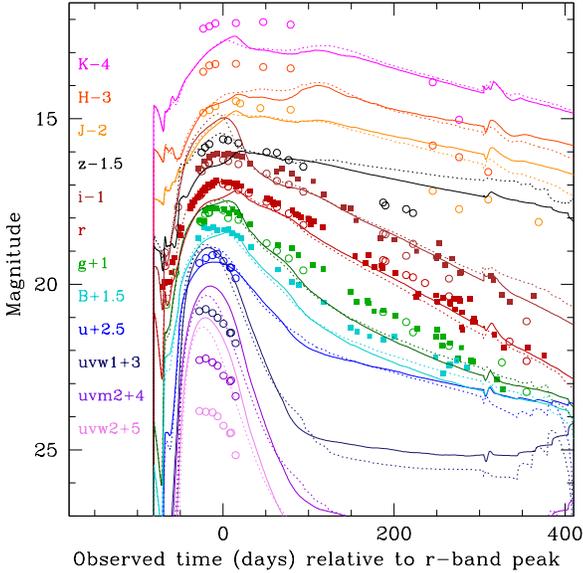}
\caption{Multicolor light curves simulation for PTF12dam in the model {\sc M80R165E20(CSM47)} and comparison with observations (filled 
squares from \citet{Vreeswijk2016}, open circles from \citet{Nicholl2013} and \citet{Chen2015}). Dotted line denotes standard resolution simulation (N$_{\rm r}$=289, N$_{\rm f}$=100), solid line - high-resolution simulation (N$_{\rm r}$=483, N$_{\rm f}$=500). Explosion time t$_{\rm 0}$=-82d.}
\label{mlcBest}
\end{figure}

\begin{figure}
\includegraphics[width=80mm]{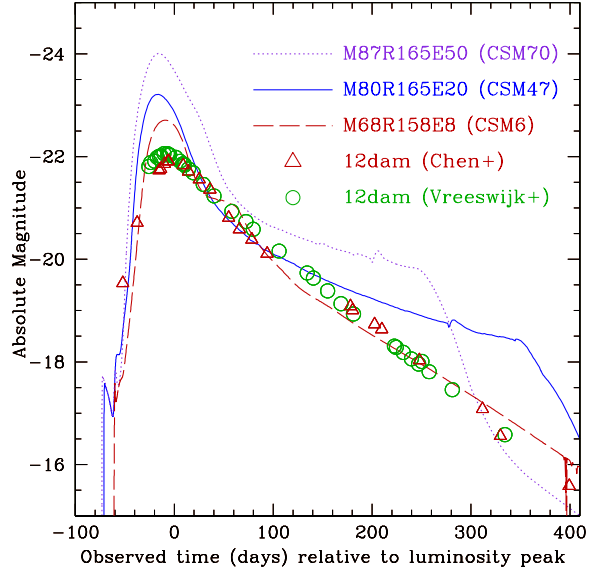}
\caption{Quasi-bolometric light curve simulations of PTF12dam for a number of interaction models with different explosion energy: $E_{51}$=50 (CSM70), $E_{51}$=20 (CSM37), $E_{51}$=8 (CSM6). The open triangles show the PTF12dam quasi-bolometric light curve as inferred 
by \citet{Vreeswijk2016}, open circles - by \citet{Chen2015} and updated by Chen et al. (2016, Erratum, in prep.).  }
\label{blcBest}
\end{figure}

\begin{figure}
\includegraphics[width=80mm]{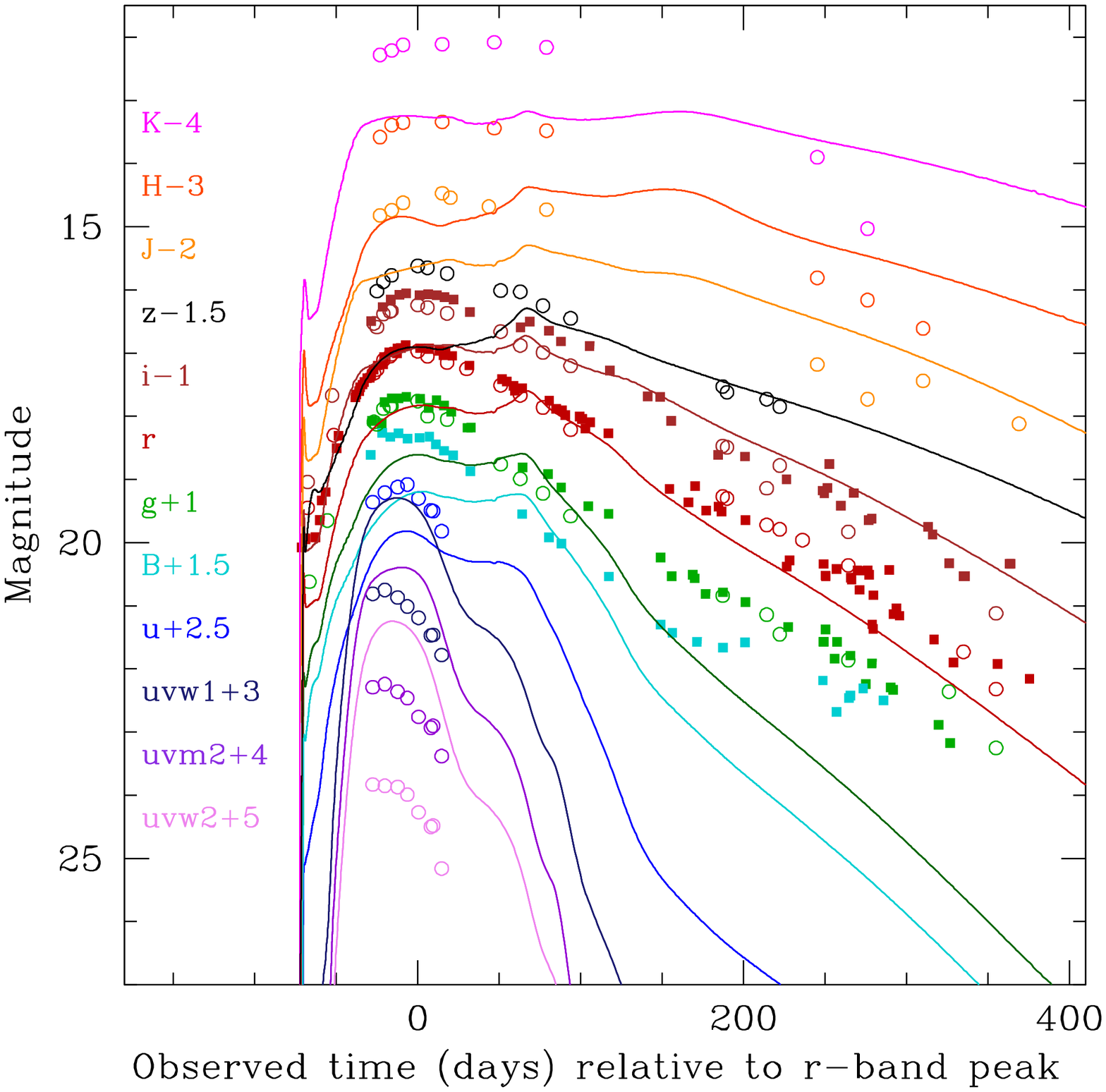}
\caption{Multicolor light curves simulation for PTF12dam in the model {\sc M68R158E8(CSM6)} and comparison with observations (filled 
squares from \citet{Vreeswijk2016}, open circles from \citet{Nicholl2013} and \citet{Chen2015}). Explosion time t$_{\rm 0}$=-73d.}
\label{mlcBestLow}
\end{figure}

\begin{figure}
\includegraphics[width=80mm]{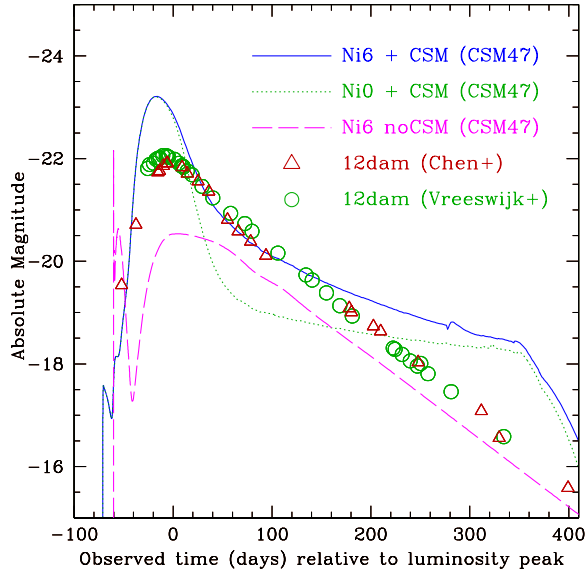}
\caption{Quasi-bolometric light curve simulations of PTF12dam in the model {\sc M80R165E20(CSM47)} varied by  the presence of CSM and $^{56}$Ni. The open triangles show the PTF12dam quasi-bolometric light curve as inferred 
by \citet{Vreeswijk2016}, open circles - by \citet{Chen2015} and updated by Chen et al. (2016, Erratum, in prep.).   }
\label{blcBestCmp}
\end{figure}

\begin{figure*}
\begin{center}
\includegraphics[width=80mm]{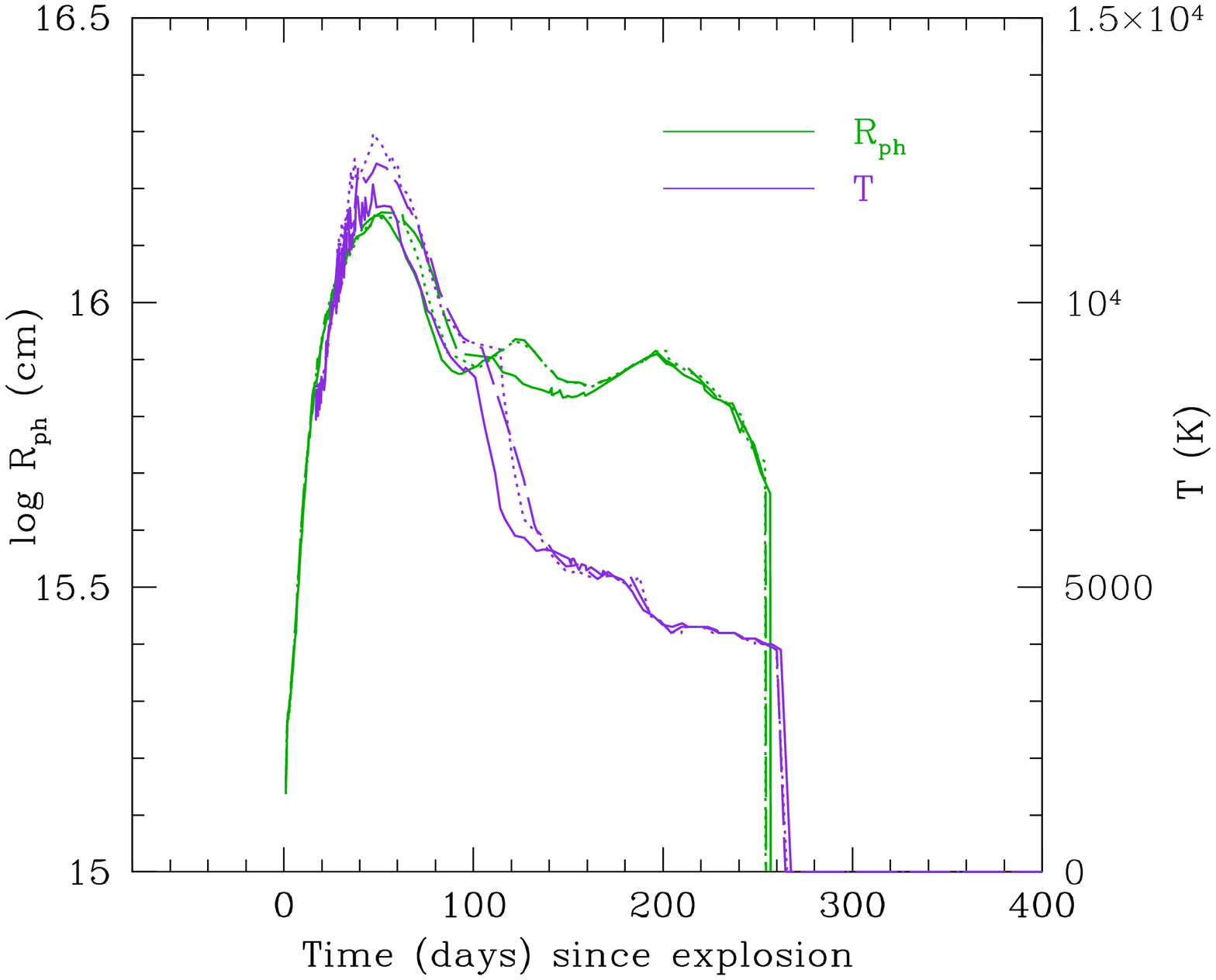}
\includegraphics[width=80mm]{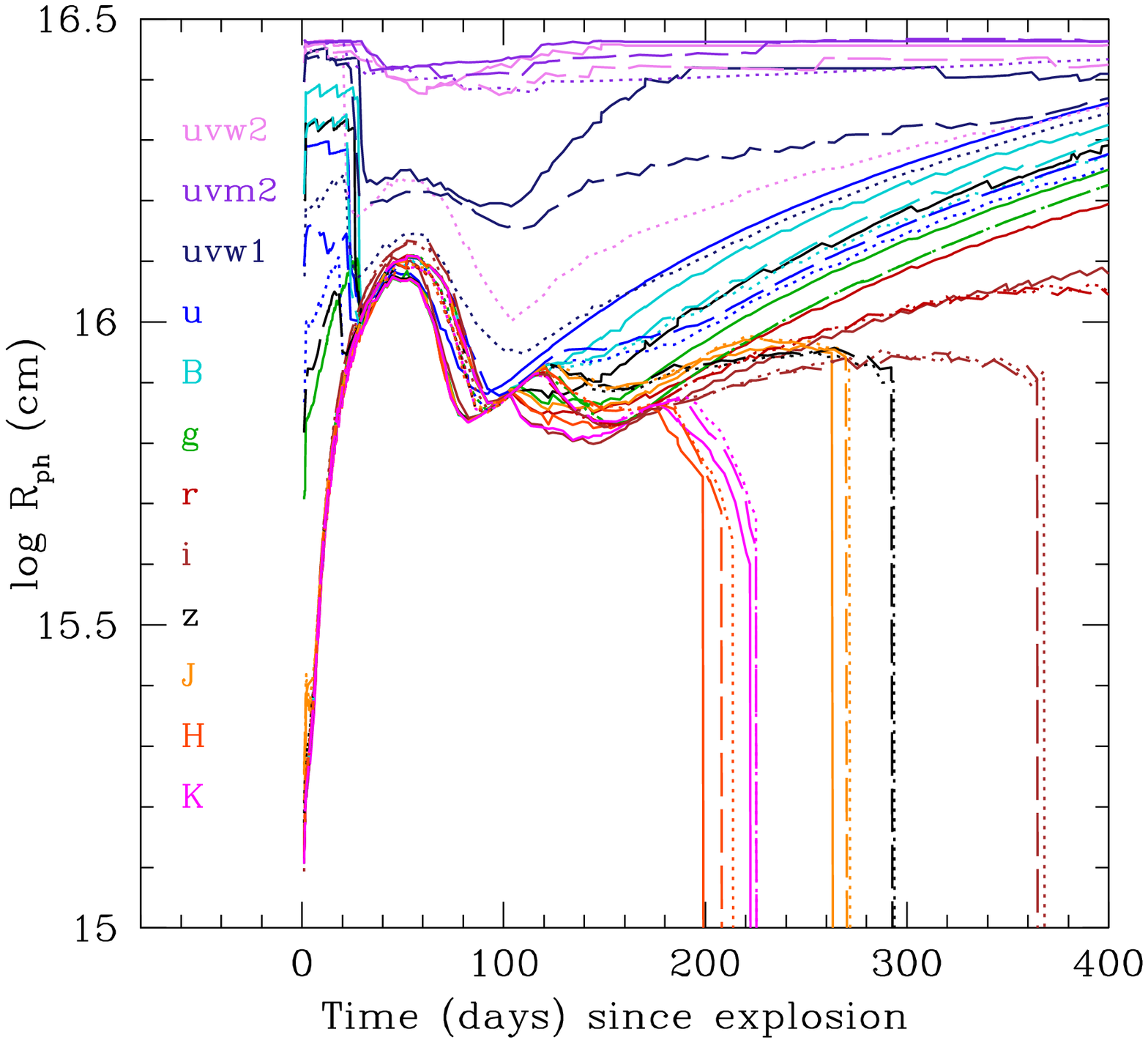}
\caption{Left panel - the photosperic radius and the temperature (Rosseland optical depth of 2/3) in the model {\sc M80R165E20(CSM47)} with solar (solid line), low Z=Z$_{\odot}$/200(dashed line), and zero metallicity (dotted line). Right panel - the photosperic radius at different broadband effective wavelengths (optical depth of 2/3). } 
\label{phRadius}
\end{center}
\end{figure*}

\begin{figure*}
\begin{center}
\includegraphics[width=80mm]{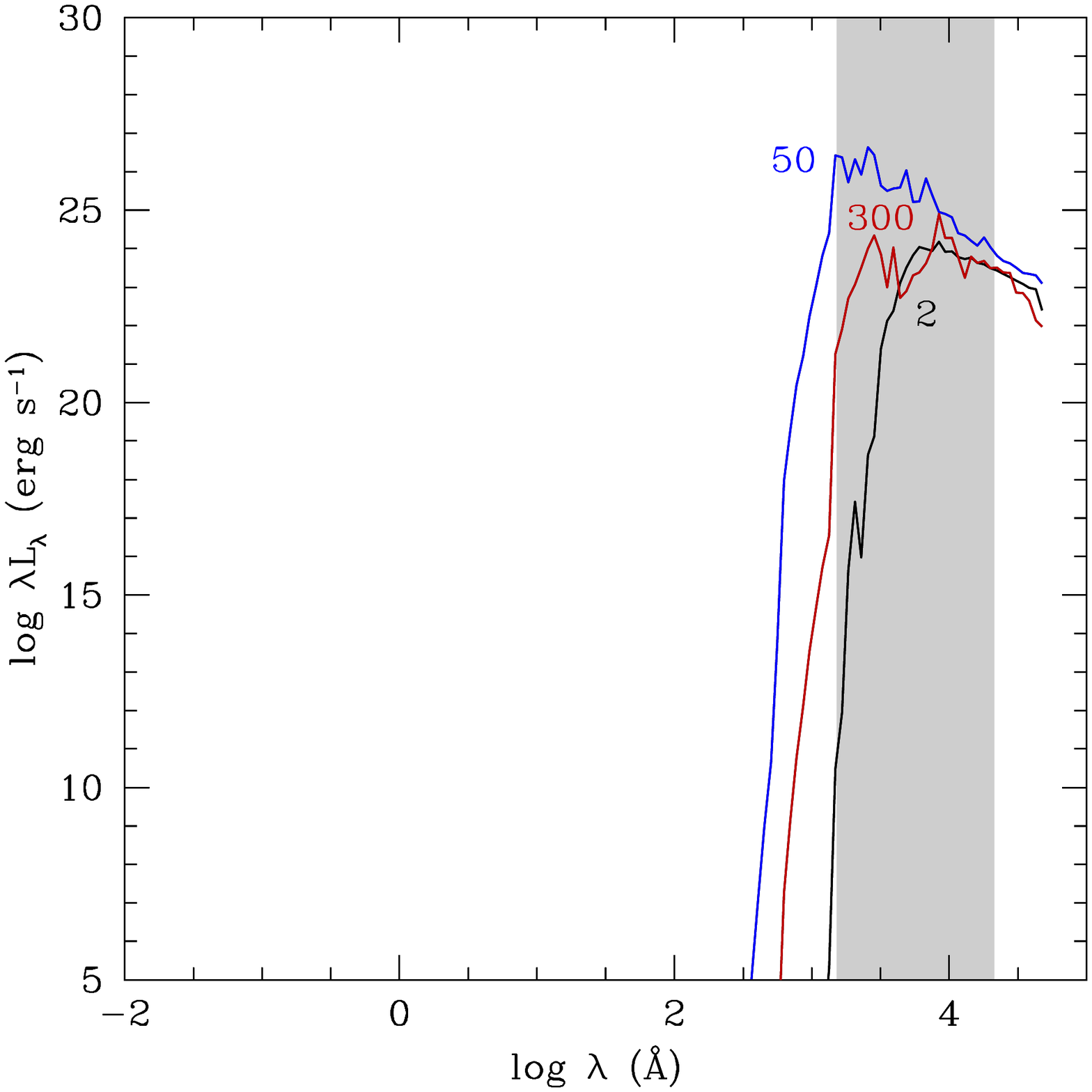}
\includegraphics[width=80mm]{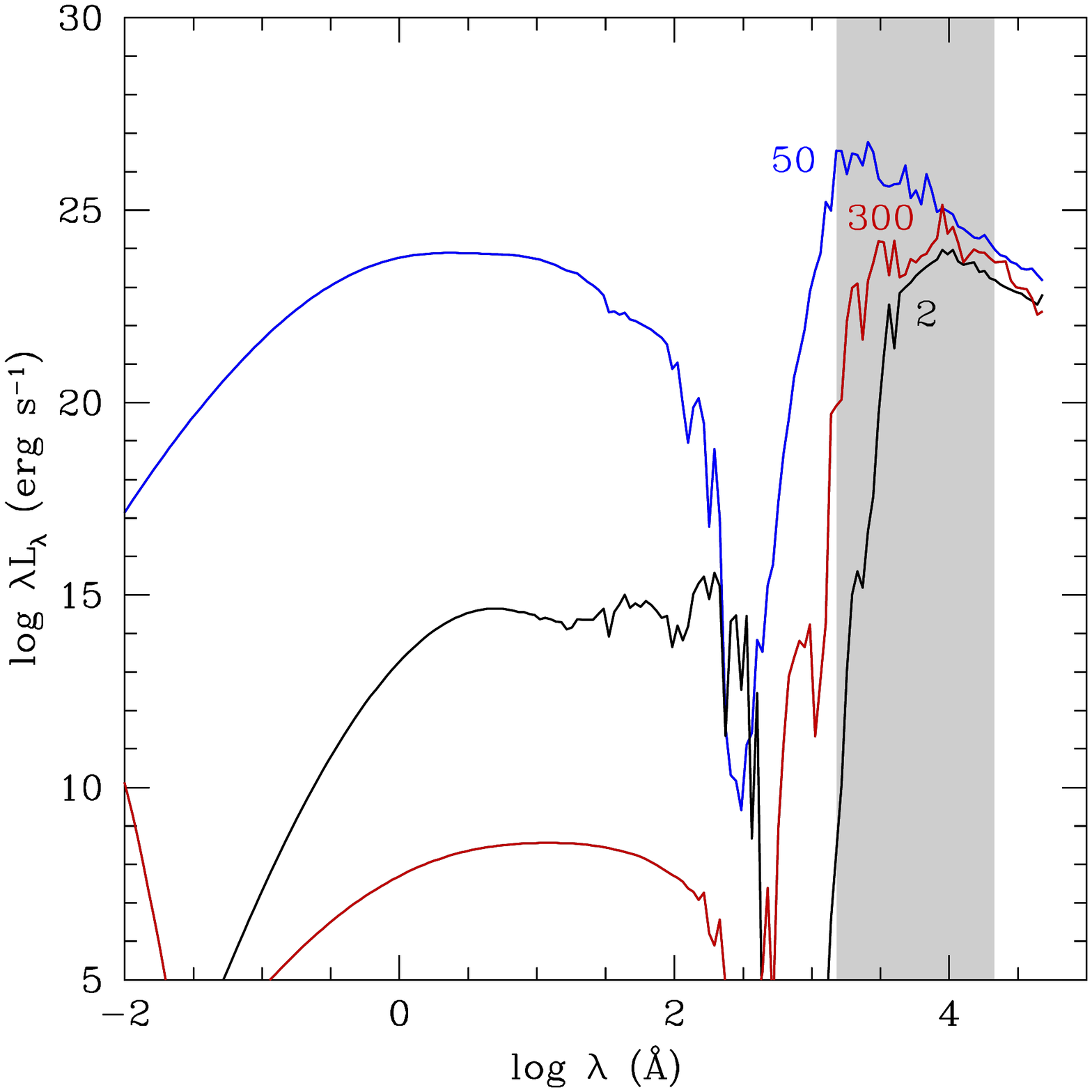}
\caption{Spectral energy distribution in the model {\sc M80R165E20(CSM47)} at different epoch with ordinary opacity (left panel) and including inner-shell photoionization (right panel). The number denotes days after explosion. Gray band corresponds to the wavelength range for calculation of quasi-bolometric light curve. } 
\label{spectra}
\end{center}
\end{figure*}

The quasi-bolometric light curve of the best-fit model is shown in Figure \ref{blcBest}. Here we note that quasi-bolometric light curve in the literature is obtained from optical broad band light curves and special procedures to estimate the infrared and UV parts of the spectrum. The real bolometric light curve can be very different from the light curve constructed from optical bands. The modeling of multicolor light curves gives a more complete picture of the phenomenon in comparison with modeling of single bolometric light curves. In calculation of quasi-bolometric light curve zero flux is
assumed outside of the observed wavelength range (1700-23000 \AA) similar to \citet{Nicholl2013}. This wavelength range is shown in Figure \ref{spectra}. A part of the spectra is not included in this range (mostly at wavelengths $<$1700 \AA). The difference in luminosity between quasi-bolometric and bolometric light curve does not exceed 10\% for our best-fit model \smaller{M80R165E20(CSM47)}.

The light curve of our best-fit model \smaller{M80R165E20(CSM47)} is brighter near the peak as well as the late epoch of the exponential decay than observational data. This effect is even higher in the model \smaller{M87R165E50(CSM70)} with high explosion energy E$_{\rm 51}$=50. The light curves of this model is too bright to fit the observations. In contrast to high explosion energy model, model \smaller{M68R158E8(CSM6)} has quite moderate explosion energy E$_{\rm 51}$=8 and demonstrates much better fit of quasi-bolometric light curve to the observations (Figure \ref{blcBest}). But multicolor optical light curves for this model are rather faint (Figure \ref{mlcBestLow}). The peak of the light curve is formed mostly by UV radiation, and faster decline of the light curve at the tail is explained by low radius of the model $\log$ R=16.1 cm. The shock wave reaches this radius at about +150 days. After that the freely expanding ejecta leads to fast cooling of the matter. In the model \smaller{M80R165E20(CSM47)} with larger radius the shock wave remains inside CSM, the velocity of the matter is lower and the cooling is slower. The ejecta becomes optically thin at +200 days in infrared, but it is still not transparent in optical and UV bands (Figure \ref{phRadius}). 

In Figure \ref{blcBestCmp} we show what contribution to the quasi-bolometric light curve in the two component model from each component. The peak is formed mostly by the interaction (model \smaller{M80R165E20(CSM47)} with no $^{56}$Ni). Later the light curve is followed by a contribution of radioactive $^{56}$Ni decay (model \smaller{M80R165E20(CSM47)} with no CSM). The tail at $t>+200$d after the peak is formed by interaction with CSM in case the radius of the CSM is large enough ($\log$ R=16.5 cm for the \smaller{M80R165E20(CSM47)}) and the shock wave is propagating inside a dense CSM.

\subsection{Opacity}

All ordinary {\sc stella} calculations employ the assumptions
used in the code {\sc eddington} \citep{EastmanPinto1993}
for bound-free transitions, in which the effect of inner-shell photoionization is not included. We briefly investigated the influence of this effect for the model \smaller{M80R165E20(CSM47)} on spectral energy distribution (SED). The inner-shell photoionization cross-sections are based on formulae derived by \citet{Verner1993,Verner1996} and \citet{VernerYakovlev1995}. In these calculations we extended frequency grid: the number of frequency bins was doubled and the
minimum wavelength $\lambda$ was set equal to 10$^{-2} \AA$  (instead of 1 $\AA$ in the standard series).

The resulting SED for two versions of our calculations are presented in Figure \ref{spectra}. The X-ray count rate due to inner-shell photoionization reaches maximum $3\cdot 10^{-4}$ counts/s in the 0.3-10 keV Swift X-ray energy range at 34 days after the explosion that corresponds to 15 days before the luminosity maximum. More detailed analyses of the X-ray emission in interaction models will be discussed in separate paper. 

In addition to inner-shell opacity we checked the influence of inclusion of excited levels in bound-free
absorption using fitting formulae as in
code {\sc wmbasic} \citep{Pauldrach1987}, but we did not find any significant changes of the light curves.

\subsection{Metallicity of CSM}

\begin{figure}
\includegraphics[width=80mm]{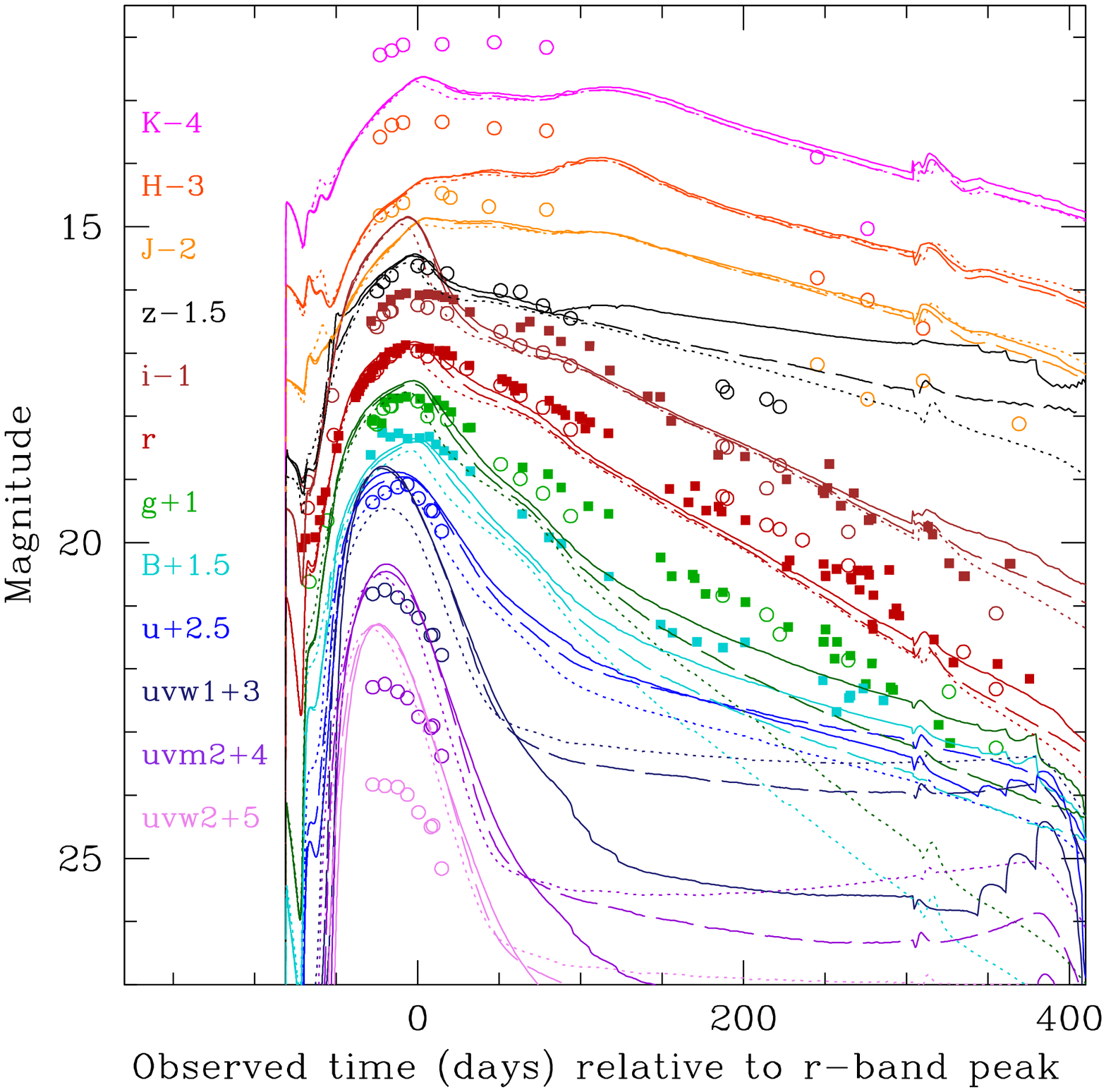}
\caption{Multicolor light curves simulation for PTF12dam in the model {\sc M80R165E20(CSM47)} with solar (solid line), low Z=Z$_{\odot}$/200(dashed line), and zero metallicity (dotted line).}
\label{lowMetal}
\end{figure}

In our calculations solar metallicity is assumed in CSM. Lower metallicity affects the tail of light curves mostly in blue and UV bands. Due to lower opacity the CSM cools down faster and the light curve decline increases (Figure \ref{lowMetal}). Another effect of lower opacity is the decrease of the radius of the photospere, especially in UV wavelengths (Figure \ref{phRadius}). The temperature of internal CSM layers is higher that leads to higher luminosity at UV wavelengths.   

\subsection{Hydrodynamics}

The evolution of hydrodynamic quantities is quite common for all of the interaction models (Figure \ref{compRhovTtaulgrAL}). The CSM is cool and transparent at the beginning.
After the shock moving with $v \sim 10,000$km/s heats the gas and reaches optical depth of stellar atmosphere $\tau \sim 50$ at about $1.5$ days after explosion, the photosphere starts to move outward from its initial radius $R_{\rm ph}=10^{15}$ cm. The expanding photosphere reaches maximum radius and luminosity in $\sim 50$ days after the explosion.
%After the shock heats the gas, the photospehere starts to move outward and finally provides the maximum luminosity. 
After the luminosity reaches the peak magnitude, the CSM and ejecta cool down and become more and more transparent, corresponding to the decline in the luminosity. Details of the process are described by \citet{Sorokina2016}.

\begin{figure}
\includegraphics[width=80mm]{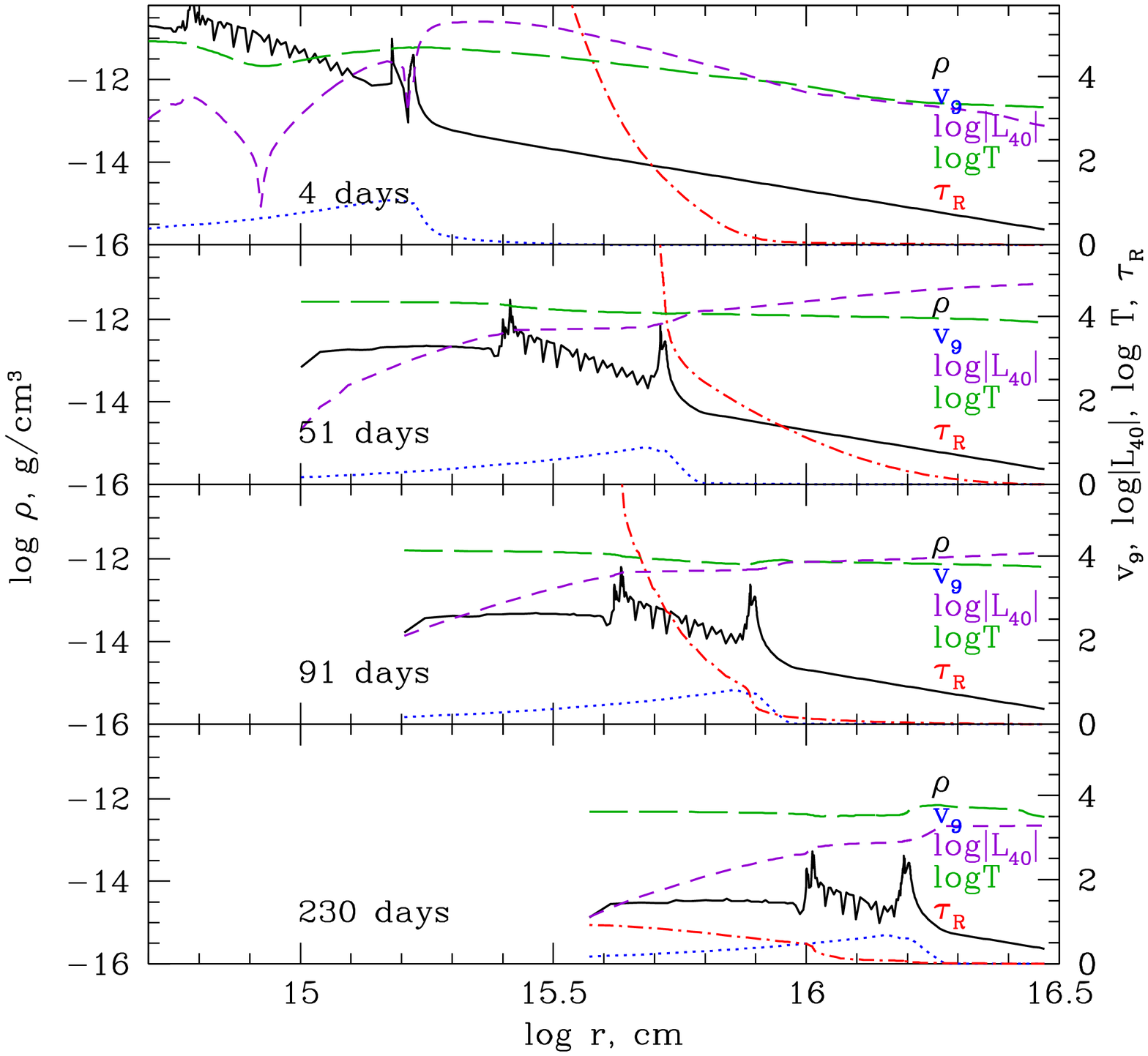}
\caption{Evolution of radial profiles of the density (solid black lines), velocity (in 10$^9$ cm$\cdot$s$^{-1}$, blue dots), matter temperature (green long dashes), and Rosseland
optical depth (red dash-dots) for the best-fit model at different time epoch after the explosion for the model {\sc M80R165E20(CSM47)}. The scale for the density is on the left Y axis, for all other quantities, on the right Y axis. 
}
\label{compRhovTtaulgrAL}
\end{figure}

\subsection{Color and effective temperatures}

The estimations of effective temperatures for PTF12dam and SN 2007bi have been performed by \citet{Nicholl2013}. They fitted blackbody curves to the optical photometric flux, and also to the continuum in the spectra. Both methods gave similar temperature estimates.

In fact the effective temperature calculated by \citet{Nicholl2013} is that, what we call color temperature $T_{\rm color}$, being the temperature of the blackbody whose spectral energy distribution fits most closely to the data. The effective temperature $T_{\rm eff}$ is defined as $T_{\rm eff}=(L/(4\pi\sigma R^2))^{1/4}$, where $\sigma$ is Stefan–Boltzmann constant, $L$ is the bolometric luminosity and $R$ is the stellar radius. Interpreting the color temperature as an effective temperature leads to non-physical formal ''radius'' \citep{Baschek1991}.

We plotted the color and effective temperature for the best-fit interaction model \smaller{M80R165E20(CSM47)} and compare them with PISN and magnetar models (Figure \ref{12damTemp}). We found that in the interaction model the color temperature near the luminosity peak is very close to the observed values.

The difference between the color and effective temperature shows that the radiation is significantly diluted. The agreement with the observations for the color temperature evolution in the interaction model \smaller{M80R165E20(CSM47)} is considerably better than for color temperatures in PISN model, published by \citet{Nicholl2013}. The color temperature for PISN models in \citet{Kozyreva2015,Kozyreva2017} is also lower than the observed temperature of PTF12dam and do not exceed 11,000 K.

In Figure \ref{12damTemp} we also show the effective temperature evolution in magnetar model, published by \citet{Nicholl2013}. It is difficult to estimate the color temperature in simple magnetar models and more detailed radiation hydro simulations are required to make a comparison.  

\begin{figure}
\includegraphics[width=80mm]{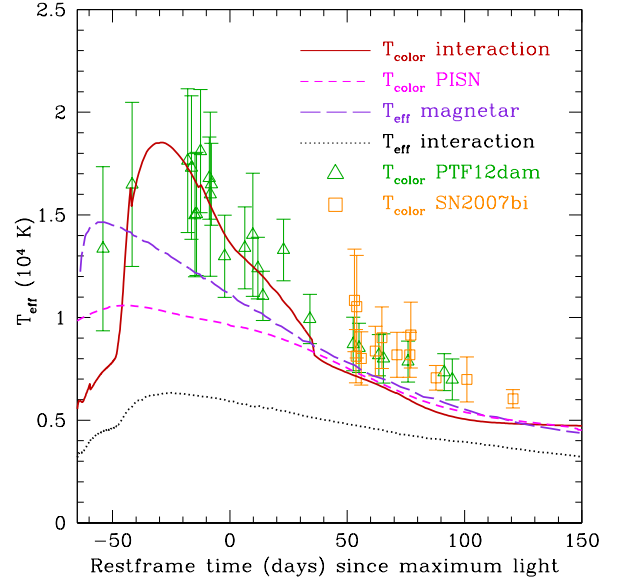}
\caption{Color temperature evolution of PTF12dam and SN 2007bi \citep{Nicholl2013}, compared with fast evolving pair-instability (short dashes) model by \citep[][P250 model]{Kozyreva2017} and interaction model \smaller{M80R165E20(CSM47)} (solid line). 
Long dashed line denotes effective temperature for magnetar-powered model \citep{Nicholl2013}.
Dotted line denotes effective temperature for interaction model.
}
\label{12damTemp}
\end{figure}

\subsection{Photosperic velocities}

\citet{Chen2015} estimated the photospheric velocities at +509 days after the luminosity peak of PTF12dam in [OI] and [CaII] lines as 4,000-6,000 km s$^{-1}$. The photospheric velocities of our best-fit model \smaller{M80R165E20(CSM47)} have a good correspondence with these values (Figure \ref{vel}). Near the luminosity peak the photospheric velocity is about 10,000 km s$^{-1}$ \citep{Nicholl2013}. Here our model has only 8,000 km s$^{-1}$, slightly lower in comparison with observed values. But the velocity near the peak is higher in the model \smaller{M58R165E20(CSM37)} with less dense CSM due to larger acceleration of the shock wave and in this model the velocity easily reaches values more than 10,000 km s$^{-1}$. Thus, the density profile of the CSM could be not so trivial as we assume and the best strategy for detailed modeling here is a construction of a detailed evolutionary model which forms the CSM. The model \smaller{M68R158E8(CSM6)} with a good fit of bolometric light curve has the photosperic velocity about 4,000 km s$^{-1}$, but this is not enough to explain observational data. High photospheric velocities $>$9,000 km s$^{-1}$ are reached in the model \smaller{M87R165E50(CSM70)} with high explosion energy E$_{\rm 51}$=50, but the luminosity in this model with M$_{\rm bol,peak}$=-24 is too high in comparison with observations.

\begin{figure}
\includegraphics[width=80mm]{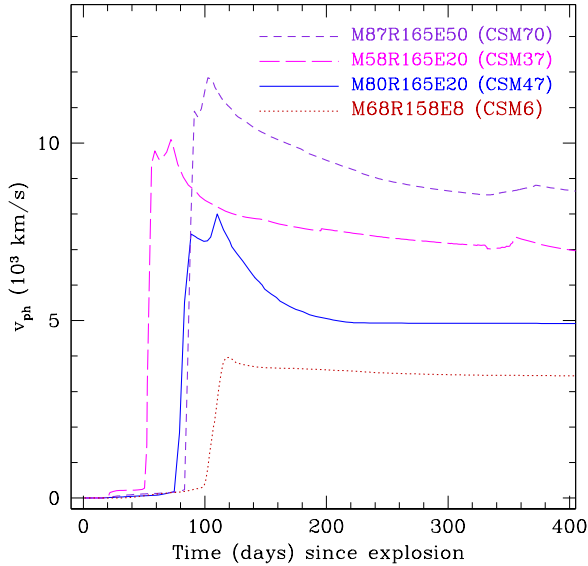}
\caption{Simulations of the photospheric velocity (at effective wavelength of B-band $\lambda_{\rm eff,B}$) for a number of interaction models with different explosion energy: $E_{51}$=50 (CSM70), $E_{51}$=20 (CSM47,CSM37), $E_{51}$=8 (CSM6). CSM37 model has lower density and mass of the CSM in comparison with CSM47 model.}  
\label{vel}
\end{figure}

\section{CONCLUSIONS}
\label{sec:conclusion}

Using detailed radiation-hydrodynamics calculations,
we constructed models for the SLSN PTF12dam in combination of the shock wave interaction with the CSM and the $^{56}$Ni radioactive decay. Our modeling shows that all of the main characteristics (the multicolor light curves, the color temperature, and the photosperic velocities) can be reproduced satisfactorily with a minimum set of model parameters. 

A large explosion energy $E_{51}$ $\sim$ 20-30 is required to produce a large mass of $^{56}$Ni as $\sim$ 6M$_{\odot}$ \citep{UmedaNomoto2008}. The explosion energy has the same magnitude as it is required in the PISN models, but in general PISN models do not easily reproduce the short rise time of the light curve, the width of the peak and the photospheric temperature. 

The most intriguing open question is the origin of
the CSM and their chemical composition, density and temperature profiles. 
%The extended shell can be formed as a result of an explosion through the pulsational pair instability mechanism proposed by \citet{HegerWoosley2002}. 
Our modeling shows that a helium CSM better describes the form of the light curve, but the presence of carbon is better for the late light curve. Most probably the CSM composition is not uniform and includes all of the elements we consider: helium, carbon, and oxigen. The origin, structure, and composition of the CSM is not clearly understood at the moment and should be clarified by evolutionary calculations including possible mass loss scenarios.

The plausible scenario is the mass ejection due to
the pulsational pair instability mechanism  proposed by \citet{Barkat1967} and taking place for the main-sequence
mass range of 80-140M$_{\odot}$ \citep{HegerWoosley2002}. The H-rich atmosphere should have probably been lost as wind in earlier stages of the evolution, and has no
effect on the light curves. During the pair instability pulsations the helium layer, possibly with some of the underlying CO is ejected, leaving a bare CO core remnant of $\sim$ 40 M$_{\odot}$, which then explodes as a CCSN.

Our modeling of PTF12dam light curves requires a massive progenitor with main-sequence mass more than 80M$_{\odot}$. It is likely that the explosion of this massive star forms a black hole rather than a neutron star \citep{Woosley2016,Makishima2016} and thus a magnetar is not formed.

%\citep{Woosley2016,Makishima2016}.

The issue of finding the most reliable scenario for PTF12dam, and SLSNe in general, requires more sophisticated models for PISN and magnetars. The combination of the modeling of multicolor light curves using radiation hydrodynamics and spectra synthesis looks like the most promising way for the identification of the scenario for SLSNe.

\acknowledgments

We thank Takashi Moriya for SN 2007bi model data,
Janet Ting-Wan Chen for PTF12dam data, and Alexandra Kozyreva for PTF12dam PISN model data and fruitful discussions. We also thank the anonymous referee for the extensive review and many useful comments. 

This research is supported by the World Premier International Research Center Initiative (WPI Initiative), MEXT, Japan, and JSPS KAKENHI Grant Numbers JP16K17658, JP26400222, JP16H02168. The work of S.B., E.S., and P.B. on the development of {\sc stella} code, in particular the implementation inner-shell
opacity for interaction models, supported by the Russian Science Foundation Grant No. 14-12-00203. P.B. is also
supported by the Swiss National Science Foundation 
(grant No. IZ73Z0\_152485 SCOPES). Numerical calculations were in part carried out on the CFCA cluster (XC30)
of National Astronomical Observatory of Japan.

\bibliographystyle{apj}
\bibliography{bibfile}

\begin{thebibliography}{33}
\providecommand\natexlab[1]{#1}
\providecommand\JournalTitle[1]{#1}

\bibitem[{{Baklanov} {et~al.}(2015){Baklanov}, {Sorokina}, \&
  {Blinnikov}}]{Baklanov2015}
{Baklanov}, P.~V., {Sorokina}, E.~I., \& {Blinnikov}, S.~I. 2015,
  \href{http://dx.doi.org/10.1134/S1063773715040027}{\JournalTitle{Astronomy
  Letters}, 41, 95}

\bibitem[{{Barkat} {et~al.}(1967){Barkat}, {Rakavy}, \& {Sack}}]{Barkat1967}
{Barkat}, Z., {Rakavy}, G., \& {Sack}, N. 1967,
  \href{http://dx.doi.org/10.1103/PhysRevLett.18.379}{\JournalTitle{Physical
  Review Letters}, 18, 379}

\bibitem[{{Baschek} {et~al.}(1991){Baschek}, {Scholz}, \&
  {Wehrse}}]{Baschek1991}
{Baschek}, B., {Scholz}, M., \& {Wehrse}, R. 1991, \JournalTitle{\aap}, 246,
  374

\bibitem[{{Blinnikov} {et~al.}(2000){Blinnikov}, {Lundqvist}, {Bartunov},
  {Nomoto}, \& {Iwamoto}}]{Blinnikov2000}
{Blinnikov}, S., {Lundqvist}, P., {Bartunov}, O., {Nomoto}, K., \& {Iwamoto},
  K. 2000, \href{http://dx.doi.org/10.1086/308588}{\JournalTitle{\apj}, 532,
  1132}

\bibitem[{{Blinnikov} {et~al.}(1998){Blinnikov}, {Eastman}, {Bartunov},
  {Popolitov}, \& {Woosley}}]{Blinnikov1998}
{Blinnikov}, S.~I., {Eastman}, R., {Bartunov}, O.~S., {Popolitov}, V.~A., \&
  {Woosley}, S.~E. 1998,
  \href{http://dx.doi.org/10.1086/305375}{\JournalTitle{\apj}, 496, 454}

\bibitem[{{Blinnikov} {et~al.}(2006){Blinnikov}, {R{\"o}pke}, {Sorokina},
  {Gieseler}, {Reinecke}, {Travaglio}, {Hillebrandt}, \&
  {Stritzinger}}]{Blinnikov2006}
{Blinnikov}, S.~I., {R{\"o}pke}, F.~K., {Sorokina}, E.~I., {et~al.} 2006,
  \href{http://dx.doi.org/10.1051/0004-6361:20054594}{\JournalTitle{\aap}, 453,
  229}

\bibitem[{{Chatzopoulos} {et~al.}(2013){Chatzopoulos}, {Wheeler}, {Vinko},
  {Horvath}, \& {Nagy}}]{Chatzopoulos2013}
{Chatzopoulos}, E., {Wheeler}, J.~C., {Vinko}, J., {Horvath}, Z.~L., \& {Nagy},
  A. 2013,
  \href{http://dx.doi.org/10.1088/0004-637X/773/1/76}{\JournalTitle{\apj}, 773,
  76}

\bibitem[{{Chen} {et~al.}(2015){Chen}, {Smartt}, {Jerkstrand}, {Nicholl},
  {Bresolin}, {Kotak}, {Polshaw}, {Rest}, {Kudritzki}, {Zheng}, {Elias-Rosa},
  {Smith}, {Inserra}, {Wright}, {Kankare}, {Kangas}, \& {Fraser}}]{Chen2015}
{Chen}, T.-W., {Smartt}, S.~J., {Jerkstrand}, A., {et~al.} 2015,
  \href{http://dx.doi.org/10.1093/mnras/stv1360}{\JournalTitle{\mnras}, 452,
  1567}

\bibitem[{{Dessart} {et~al.}(2015){Dessart}, {Audit}, \&
  {Hillier}}]{Dessart2015}
{Dessart}, L., {Audit}, E., \& {Hillier}, D.~J. 2015,
  \href{http://dx.doi.org/10.1093/mnras/stv609}{\JournalTitle{\mnras}, 449,
  4304}

\bibitem[{{Dessart} {et~al.}(2012){Dessart}, {Hillier}, {Waldman}, {Livne}, \&
  {Blondin}}]{Dessart2012}
{Dessart}, L., {Hillier}, D.~J., {Waldman}, R., {Livne}, E., \& {Blondin}, S.
  2012,
  \href{http://dx.doi.org/10.1111/j.1745-3933.2012.01329.x}{\JournalTitle{\mnras},
  426, L76}

\bibitem[{{Dessart} {et~al.}(2013){Dessart}, {Waldman}, {Livne}, {Hillier}, \&
  {Blondin}}]{Dessart2013}
{Dessart}, L., {Waldman}, R., {Livne}, E., {Hillier}, D.~J., \& {Blondin}, S.
  2013, \href{http://dx.doi.org/10.1093/mnras/sts269}{\JournalTitle{\mnras},
  428, 3227}

\bibitem[{{Eastman} \& {Pinto}(1993)}]{EastmanPinto1993}
{Eastman}, R.~G., \& {Pinto}, P.~A. 1993,
  \href{http://dx.doi.org/10.1086/172957}{\JournalTitle{\apj}, 412, 731}

\bibitem[{{Gal-Yam} {et~al.}(2009){Gal-Yam}, {Mazzali}, {Ofek}, {Nugent},
  {Kulkarni}, {Kasliwal}, {Quimby}, {Filippenko}, {Cenko}, {Chornock},
  {Waldman}, {Kasen}, {Sullivan}, {Beshore}, {Drake}, {Thomas}, {Bloom},
  {Poznanski}, {Miller}, {Foley}, {Silverman}, {Arcavi}, {Ellis}, \&
  {Deng}}]{GalYam2009}
{Gal-Yam}, A., {Mazzali}, P., {Ofek}, E.~O., {et~al.} 2009,
  \href{http://dx.doi.org/10.1038/nature08579}{\JournalTitle{\nat}, 462, 624}

\bibitem[{{Heger} \& {Woosley}(2002)}]{HegerWoosley2002}
{Heger}, A., \& {Woosley}, S.~E. 2002,
  \href{http://dx.doi.org/10.1086/338487}{\JournalTitle{\apj}, 567, 532}

\bibitem[{{Kozyreva} \& {Blinnikov}(2015)}]{Kozyreva2015}
{Kozyreva}, A., \& {Blinnikov}, S. 2015,
  \href{http://dx.doi.org/10.1093/mnras/stv2287}{\JournalTitle{\mnras}, 454,
  4357}

\bibitem[{{Kozyreva} {et~al.}(2014){Kozyreva}, {Blinnikov}, {Langer}, \&
  {Yoon}}]{Kozyreva2014}
{Kozyreva}, A., {Blinnikov}, S., {Langer}, N., \& {Yoon}, S.-C. 2014,
  \href{http://dx.doi.org/10.1051/0004-6361/201423447}{\JournalTitle{\aap},
  565, A70}

\bibitem[{{Kozyreva} {et~al.}(2017){Kozyreva}, {Gilmer}, {Hirschi},
  {Fr{\"o}hlich}, {Blinnikov}, {Wollaeger}, {Noebauer}, {van Rossum}, {Heger},
  {Even}, {Waldman}, {Tolstov}, {Chatzopoulos}, \& {Sorokina}}]{Kozyreva2017}
{Kozyreva}, A., {Gilmer}, M., {Hirschi}, R., {et~al.} 2017,
  \href{http://dx.doi.org/10.1093/mnras/stw2562}{\JournalTitle{\mnras}, 464,
  2854}

\bibitem[{{Kurucz} \& {Bell}(1995)}]{KuruczBell1995}
{Kurucz}, R.~L., \& {Bell}, B. 1995, {Atomic line list}

\bibitem[{{Makishima}(2016)}]{Makishima2016}
{Makishima}, K. 2016,
  \href{http://dx.doi.org/10.2183/pjab.92.135}{\JournalTitle{Proceeding of the
  Japan Academy, Series B}, 92, 135}

\bibitem[{{Moriya} {et~al.}(2010){Moriya}, {Tominaga}, {Tanaka}, {Maeda}, \&
  {Nomoto}}]{Moriya2010}
{Moriya}, T., {Tominaga}, N., {Tanaka}, M., {Maeda}, K., \& {Nomoto}, K. 2010,
  \href{http://dx.doi.org/10.1088/2041-8205/717/2/L83}{\JournalTitle{\apjl},
  717, L83}

\bibitem[{{Nicholl} {et~al.}(2013){Nicholl}, {Smartt}, {Jerkstrand}, {Inserra},
  {McCrum}, {Kotak}, {Fraser}, {Wright}, {Chen}, {Smith}, {Young}, {Sim},
  {Valenti}, {Howell}, {Bresolin}, {Kudritzki}, {Tonry}, {Huber}, {Rest},
  {Pastorello}, {Tomasella}, {Cappellaro}, {Benetti}, {Mattila}, {Kankare},
  {Kangas}, {Leloudas}, {Sollerman}, {Taddia}, {Berger}, {Chornock}, {Narayan},
  {Stubbs}, {Foley}, {Lunnan}, {Soderberg}, {Sanders}, {Milisavljevic},
  {Margutti}, {Kirshner}, {Elias-Rosa}, {Morales-Garoffolo}, {Taubenberger},
  {Botticella}, {Gezari}, {Urata}, {Rodney}, {Riess}, {Scolnic}, {Wood-Vasey},
  {Burgett}, {Chambers}, {Flewelling}, {Magnier}, {Kaiser}, {Metcalfe},
  {Morgan}, {Price}, {Sweeney}, \& {Waters}}]{Nicholl2013}
{Nicholl}, M., {Smartt}, S.~J., {Jerkstrand}, A., {et~al.} 2013,
  \href{http://dx.doi.org/10.1038/nature12569}{\JournalTitle{\nat}, 502, 346}

\bibitem[{{Ohkubo} {et~al.}(2009){Ohkubo}, {Nomoto}, {Umeda}, {Yoshida}, \&
  {Tsuruta}}]{Ohkubo2009}
{Ohkubo}, T., {Nomoto}, K., {Umeda}, H., {Yoshida}, N., \& {Tsuruta}, S. 2009,
  \href{http://dx.doi.org/10.1088/0004-637X/706/2/1184}{\JournalTitle{\apj},
  706, 1184}

\bibitem[{{Pauldrach}(1987)}]{Pauldrach1987}
{Pauldrach}, A. 1987, \JournalTitle{\aap}, 183, 295

\bibitem[{{Quimby}(2014)}]{Quimby2014}
{Quimby}, R.~M. 2014, \href{http://dx.doi.org/10.1017/S1743921313009253}{in IAU
  Symposium, Vol. 296, Supernova Environmental Impacts, ed. A.~{Ray} \& R.~A.
  {McCray}}, 68

\bibitem[{{Sorokina} {et~al.}(2016){Sorokina}, {Blinnikov}, {Nomoto}, {Quimby},
  \& {Tolstov}}]{Sorokina2016}
{Sorokina}, E., {Blinnikov}, S., {Nomoto}, K., {Quimby}, R., \& {Tolstov}, A.
  2016,
  \href{http://dx.doi.org/10.3847/0004-637X/829/1/17}{\JournalTitle{\apj}, 829,
  17}

\bibitem[{{Umeda} \& {Nomoto}(2008)}]{UmedaNomoto2008}
{Umeda}, H., \& {Nomoto}, K. 2008,
  \href{http://dx.doi.org/10.1086/524767}{\JournalTitle{\apj}, 673, 1014}

\bibitem[{{Verner} {et~al.}(1996){Verner}, {Ferland}, {Korista}, \&
  {Yakovlev}}]{Verner1996}
{Verner}, D.~A., {Ferland}, G.~J., {Korista}, K.~T., \& {Yakovlev}, D.~G. 1996,
  \href{http://dx.doi.org/10.1086/177435}{\JournalTitle{\apj}, 465, 487}

\bibitem[{{Verner} \& {Yakovlev}(1995)}]{VernerYakovlev1995}
{Verner}, D.~A., \& {Yakovlev}, D.~G. 1995, \JournalTitle{\aaps}, 109

\bibitem[{{Verner} {et~al.}(1993){Verner}, {Yakovlev}, {Band}, \&
  {Trzhaskovskaya}}]{Verner1993}
{Verner}, D.~A., {Yakovlev}, D.~G., {Band}, I.~M., \& {Trzhaskovskaya}, M.~B.
  1993, \href{http://dx.doi.org/10.1006/adnd.1993.1022}{\JournalTitle{Atomic
  Data and Nuclear Data Tables}, 55, 233}

\bibitem[{{Vreeswijk} {et~al.}(2016){Vreeswijk}, {Leloudas}, {Gal-Yam}, {De
  Cia}, {Perley}, {Quimby}, {Waldman}, {Sullivan}, {Yan}, {Ofek}, {Fremling},
  {Taddia}, {Sollerman}, {Valenti}, {Arcavi}, {Howell}, {Filippenko}, {Cenko},
  {Yaron}, {Kasliwal}, {Cao}, {Ben-Ami}, {Horesh}, {Rubin}, {Lunnan}, {Nugent},
  {Laher}, {Rebbapragada}, {Wo{\'z}niak}, \& {Kulkarni}}]{Vreeswijk2016}
{Vreeswijk}, P.~M., {Leloudas}, G., {Gal-Yam}, A., {et~al.} 2016,
  \JournalTitle{ArXiv e-prints},
  \href{http://arxiv.org/abs/1609.08145}{{\sffamily arXiv:1609.08145
  [astro-ph.HE]}}

\bibitem[{{Woosley}(2016)}]{Woosley2016}
{Woosley}, S.~E. 2016,
  \href{http://dx.doi.org/10.3847/2041-8205/824/1/L10}{\JournalTitle{\apjl},
  824, L10}

\bibitem[{{Woosley} {et~al.}(2007){Woosley}, {Blinnikov}, \&
  {Heger}}]{Woosley2007}
{Woosley}, S.~E., {Blinnikov}, S., \& {Heger}, A. 2007,
  \href{http://dx.doi.org/10.1038/nature06333}{\JournalTitle{\nat}, 450, 390}

\bibitem[{{Yoshida} {et~al.}(2016){Yoshida}, {Umeda}, {Maeda}, \&
  {Ishii}}]{Yoshida2016}
{Yoshida}, T., {Umeda}, H., {Maeda}, K., \& {Ishii}, T. 2016,
  \href{http://dx.doi.org/10.1093/mnras/stv3002}{\JournalTitle{\mnras}, 457,
  351}

\end{thebibliography}

\appendix

\section{PTF12dam interaction models}
\label{appModels}

In table \ref{modelTable2} we present all of the models which has been considered in our modeling of the light curves for PTF12dam. The parameters of the models were selected to take into account the constraints on the explosion energy, the mass of $^{56}$Ni, chemical composition and to vary the shape of the light curves around best-fit values.

%============================= TableBegin
\LongTables
%\begin{longtable*}{ p{.10\textwidth}  p{.10\textwidth}   p{.%10\textwidth}   p{.10\textwidth}  p{.10\textwidth}  p{.10\textwidth}  %p{.10\textwidth}  p{.10\textwidth} p{.10\textwidth}} 
\begin{deluxetable*}{lcccccccc}
\tablecaption{Model parameters \label{modelTable2}}
%\tablecolumns{5}
\tablewidth{0pt}
\tablehead{
\colhead{Model} & 
\colhead{$\log R$} & 
\colhead{$M$} &
\colhead{$p$} &
\colhead{$E$} &
\colhead{$M$($^{56}$Ni)} & 
\colhead{$T$} &  
\colhead{Composition} &
\colhead{$N_{r}$} 
\\
\colhead{} & 
\colhead{(cm)} & 
\colhead{($M_{\odot}$)} & 
\colhead{} &
\colhead{($E_{51}$)} &
\colhead{($M_{\odot}$)} & 
\colhead{(K)} & 
\colhead{} & 
\colhead{} 
}  

\\ 
CSM1 & 14.4 & 42 & 2.0 & 23 & 6 & 2500 & HeC & 289
\\ 
CSM2 & 15.1 & 76 & 3.5 & 20 & 6 & 1000 & He & 289
\\ 
CSM3 & 15.1 & 76 & 3.5 & 30 & 6 & 1000 & He & 289
\\ 
CSM4 & 15.3 & 54 & 2.5 & 10 & 6 & 1000 & He & 289
\\ 
CSM5 & 15.8 & 62 & 3.0 & 10 & 6 & 1000 & He & 289
\\ 
CSM6 & 15.8 & 68 & 2.5 & 10 & 6 & 1000 & He & 289
\\ 
CSM7 & 15.8 & 68 & 2.5 & 20 & 6 & 1000 & He & 289
\\ 
CSM8 & 15.8 & 68 & 2.5 & 30 & 6 & 1000 & He & 289
\\ 
CSM9 & 15.8 & 90 & 1.8 & 5 & 1 & 1000 & He & 289
\\ 
CSM10 & 15.8 & 90 & 1.8 & 5 & 6 & 1000 & He & 289
\\ 
CSM11 & 15.8 & 90 & 1.8 & 10 & 3 & 1000 & He & 289
\\ 
CSM12 & 15.8 & 90 & 1.8 & 10 & 6 & 1000 & He & 289
\\ 
CSM13 & 15.8 & 90 & 1.8 & 10 & 6 & 1000 & He & 289
\\ 
CSM14 & 15.8 & 90 & 1.8 & 10 & 6 & 2000 & He & 289
\\ 
CSM15 & 15.8 & 90 & 1.8 & 30 & 6 & 1000 & He & 289
\\ 
CSM16 & 15.8 & 93 & 3.5 & 20 & 6 & 1000 & He & 289
\\ 
CSM17 & 15.8 & 117 & 2.5 & 10 & 6 & 1000 & He & 289
\\ 
CSM18 & 15.8 & 117 & 2.5 & 20 & 6 & 1000 & He & 289
\\ 
CSM19 & 16.0 & 66 & 2.0 & 30 & 6 & 2500 & OC & 192
\\ 
CSM20 & 16.1 & 39 & 2.0 & 15 & 6 & 2500 & OC & 192
\\ 
CSM21 & 16.1 & 39 & 2.0 & 20 & 6 & 2500 & OC & 192
\\ 
CSM22 & 16.1 & 55 & 2.5 & 20 & 6 & 1000 & He & 289
\\ 
CSM23 & 16.1 & 55 & 2.5 & 30 & 6 & 1000 & He & 289
\\ 
CSM24 & 16.1 & 61 & 2.0 & 30 & 6 & 2500 & OC & 289
\\ 
CSM25 & 16.1 & 61 & 3.5 & 20 & 6 & 1000 & He & 289
\\ 
CSM26 & 16.1 & 61 & 3.5 & 30 & 6 & 1000 & He & 289
\\ 
CSM27 & 16.1 & 67 & 2.5 & 20 & 6 & 1000 & He & 289
\\ 
CSM28 & 16.1 & 67 & 2.5 & 30 & 6 & 1000 & He & 289
\\ 
CSM29 & 16.1 & 79 & 3.5 & 20 & 6 & 1000 & He & 289
\\ 
CSM30 & 16.1 & 85 & 2.0 & 30 & 3 & 2500 & OC & 192
\\ 
CSM31 & 16.1 & 85 & 2.0 & 30 & 6 & 2500 & OC & 192
\\ 
CSM32 & 16.1 & 97 & 1.5 & 30 & 6 & 2500 & OC & 192
\\ 
CSM33 & 16.1 & 97 & 1.8 & 30 & 6 & 2500 & OC & 192
\\ 
CSM34 & 16.2 & 67 & 3.0 & 10 & 6 & 1000 & He & 289
\\ 
CSM35 & 16.4 & 74 & 2.0 & 25 & 6 & 2500 & HeC2 & 289
\\ 
CSM36 & 16.4 & 107 & 1.8 & 25 & 6 & 2500 & He & 289
\\ 
CSM37 & 16.5 & 58 & 2.0 & 23 & 6 & 2500 & HeC & 289
\\ 
CSM38 & 16.5 & 80 & 2.0 & 23 & 6 & 2500 & HeC & 192
\\ 
CSM39 & 16.5 & 80 & 2.0 & 20 & 6 & 2500 & HeC & 289
\\ 
CSM40 & 16.5 & 80 & 2.0 & 23 & 0 & 2500 & HeC & 289
\\ 
CSM41 & 16.5 & 80 & 2.0 & 23 & 1 & 2500 & HeC & 289
\\ 
CSM42 & 16.5 & 80 & 2.0 & 23 & 2 & 2500 & HeC & 289
\\ 
CSM43 & 16.5 & 80 & 2.0 & 23 & 4 & 2500 & HeC & 289
\\ 
CSM44 & 16.5 & 80 & 2.0 & 23 & 6 & 2500 & CO & 289
\\ 
CSM45 & 16.5 & 80 & 2.0 & 23 & 6 & 2500 & CHeO & 289
\\ 
CSM46 & 16.5 & 80 & 2.0 & 23 & 6 & 2500 & HeCO2 & 289
\\ 
CSM47 & 16.5 & 80 & 2.0 & 23 & 6 & 2500 & HeC & 289
\\ 
CSM48 & 16.5 & 80 & 2.0 & 25 & 6 & 2500 & HeC & 289
\\ 
CSM49 & 16.5 & 80 & 2.0 & 23 & 6 & 2500 & HeC & 483
\\ 
CSM50 & 16.5 & 80 & 2.0 & 23 & 6 & 2500 & HeC & 144
\\ 
CSM51 & 16.5 & 80 & 2.0 & 23 & 6 & 2500 & HeC & 700
\\ 
CSM52 & 16.5 & 80 & 2.0 & 23 & 6 & 2500 & HeC & 900
\\ 
CSM53 & 16.5 & 56 & 2.5 & 20 & 6 & 2500 & He & 289
\\ 
CSM54 & 16.5 & 61 & 2.0 & 30 & 6 & 2500 & OC & 289
\\ 
CSM55 & 16.5 & 63 & 2.5 & 10 & 6 & 1000 & He & 289
\\ 
CSM56 & 16.5 & 63 & 2.5 & 20 & 6 & 1000 & He & 289
\\ 
CSM57 & 16.5 & 63 & 2.5 & 30 & 6 & 1000 & He & 289
\\ 
CSM58 & 16.5 & 65 & 2.0 & 20 & 6 & 2500 & He & 289
\\ 
CSM59 & 16.5 & 72 & 3.0 & 30 & 6 & 2500 & OC & 289
\\ 
CSM60 & 16.5 & 83 & 2.5 & 20 & 6 & 1000 & He & 289
\\ 
CSM61 & 16.5 & 83 & 2.5 & 30 & 6 & 1000 & He & 289
\\ 
CSM62 & 16.5 & 87 & 2.0 & 20 & 6 & 2500 & He & 289
\\ 
CSM63 & 16.5 & 87 & 2.0 & 20 & 6 & 2500 & HeCO & 289
\\ 
CSM64 & 16.5 & 87 & 2.0 & 25 & 6 & 2500 & He & 289
\\ 
CSM65 & 16.5 & 87 & 2.0 & 25 & 6 & 2500 & HeC & 289
\\ 
CSM66 & 16.5 & 87 & 2.0 & 25 & 6 & 2500 & HeC2 & 289
\\ 
CSM67 & 16.5 & 87 & 2.0 & 25 & 6 & 2500 & He & 289
\\ 
CSM68 & 16.5 & 87 & 2.0 & 30 & 6 & 2500 & He & 289
\\ 
CSM69 & 16.5 & 87 & 2.0 & 30 & 6 & 2500 & OC & 289
\\ 
CSM70 & 16.5 & 87 & 2.0 & 60 & 6 & 2500 & He & 289
\\ 
CSM71 & 16.5 & 93 & 1.7 & 20 & 6 & 2500 & He & 289
\\ 
CSM72 & 16.5 & 93 & 1.7 & 25 & 6 & 2500 & He & 289
\\ 
CSM73 & 16.5 & 109 & 2.0 & 30 & 6 & 2500 & OC & 289
\\ 
CSM74 & 16.5 & 119 & 1.8 & 20 & 6 & 2500 & He & 289
\\ 
CSM75 & 16.7 & 46 & 3.0 & 20 & 6 & 1000 & He & 289
\\ 
CSM76 & 16.7 & 53 & 3.0 & 20 & 6 & 1000 & He & 289
\\ 
CSM77 & 16.7 & 62 & 3.0 & 20 & 6 & 1000 & He & 289
\\ 
CSM78 & 16.7 & 95 & 2.5 & 20 & 6 & 1000 & He & 289
\\ 
CSM79 & 16.8 & 179 & 2.0 & 30 & 6 & 2500 & OC & 192
\\ 
CSM80 & 16.8 & 95 & 2.0 & 30 & 6 & 2500 & OC & 289
\\ 
CSM81 & 17.1 & 114 & 2.0 & 30 & 6 & 2500 & OC & 289
\\ 
CSM82 & 17.1 & 114 & 2.0 & 30 & 6 & 2500 & OC & 483
\\ 
CSM83 & 17.1 & 130 & 2.0 & 30 & 6 & 1000 & OC & 289
\\ 
CSM84 & 17.1 & 130 & 2.0 & 30 & 6 & 1500 & OC & 289
\\ 
CSM85 & 17.1 & 130 & 2.0 & 30 & 6 & 2500 & OC & 289
\\ 
CSM86 & 17.1 & 130 & 2.0 & 30 & 6 & 2500 & He & 289
\\ 
CSM87 & 17.3 & 166 & 2.0 & 30 & 6 & 2500 & OC & 289
\\ 
CSM88 & 17.3 & 459 & 2.0 & 30 & 6 & 2500 & OC & 192
\\ 
CSM89 & 17.4 & 201 & 2.0 & 30 & 6 & 2500 & He & 289
\\ 
CSM90 & 17.4 & 201 & 2.0 & 30 & 6 & 2500 & OC & 289
\\

\vspace{-0.2cm}

%\enddata
%\vspace{-0.8cm}
\tablecomments{The columns show the indexed name of the model, the outer radius of the CSM, the total mass $M=M_{\rm ej}+M_{\rm CSM}$, the index of the power-law CSM density profile, the deposited energy (kinetic energy of the ejecta in 1 day after the explosion), the mass of $^{56}$Ni, the temperature of the CSM, the composition of the wind, and the number of radius zones. The composition corresponds to the following mass fractions: OC(O:C=4:1), CO(C:O=4:1), HeC(He:C=6:1), HeC2(He:C=4:1), HeCO(He:C:O=4:1:1), HeCO2(He:C:O=4:2:1), HeCO3(C:He:O=2:1:1). The mass of SN ejecta in all the models $M_{\rm ej}$=40$M_{\odot}$.}
%\end{longtable*}
\end{deluxetable*}
%============================= TableEnd

\end{document}